\magnification \magstep1
\raggedbottom
\openup 2\jot
\voffset6truemm
\def\II{{\rm 1\!\hskip-1pt I}}
\def\cstok#1{\leavevmode\thinspace\hbox{\vrule\vtop{\vbox{\hrule\kern1pt
\hbox{\vphantom{\tt/}\thinspace{\tt#1}\thinspace}}
\kern1pt\hrule}\vrule}\thinspace}
\centerline {\bf NON-LOCAL BOUNDARY CONDITIONS IN}
\centerline {\bf EUCLIDEAN QUANTUM GRAVITY}
\vskip 0.3cm
\leftline {Giampiero Esposito}
\vskip 0.3cm
\noindent
{\it INFN, Sezione di Napoli, Mostra d'Oltremare Padiglione 20,
80125 Napoli, Italy}
\vskip 0.3cm
\noindent
{\it Universit\`a di Napoli Federico II, Dipartimento di Scienze
Fisiche, Complesso Universitario di Monte S. Angelo, Via Cintia,
Edificio G, 80126 Napoli, Italy}
\vskip 1cm
\noindent
{\bf Abstract}.
Non-local boundary conditions for Euclidean quantum gravity are
proposed, consisting of an integro-differential boundary
operator acting on metric perturbations. In this case, the operator
$P$ on metric perturbations is of Laplace type, subject to
non-local boundary conditions; by contrast, its adjoint is the
sum of a Laplacian and of a singular Green operator, subject to
local boundary conditions. Self-adjointness of the boundary value
problem is correctly formulated by looking at Dirichlet-type and
Neumann-type realizations of the operator $P$, following 
recent results in the literature. The set of non-local boundary
conditions for perturbative modes of the gravitational field is
written in general form on the Euclidean 4-ball. For a particular
choice of the non-local boundary operator, explicit formulae for
the boundary value problem are obtained in terms of a finite number
of unknown functions, but subject to some consistency 
conditions. Among the
related issues, the problem arises of whether non-local symmetries
exist in Euclidean quantum gravity.
\vskip 100cm
\leftline {\bf 1. Introduction}
\vskip 0.3cm
\noindent
The last decade of efforts on the problem of boundary conditions
in (one-loop) Euclidean quantum gravity has focused on a 
{\it local formulation}, by trying to satisfy the following 
requirements:
\vskip 0.3cm
\noindent
(i) Local nature of the boundary operators [1--5].
\vskip 0.3cm
\noindent
(ii) Operator on metric perturbations, say $P$, and ghost operator,
say $Q$, of Laplace type [5].
\vskip 0.3cm
\noindent
(iii) Symmetry, and, possibly, (essential) self-adjointness of the
differential operators $P$ and $Q$ [4, 5].
\vskip 0.3cm
\noindent
(iv) Strong ellipticity of the boundary value problems obtained
from the operators $P$ and $Q$, 
with local boundary operators $B_{1}$ and
$B_{2}$, respectively [5, 6].
\vskip 0.3cm
\noindent
(v) Gauge- and BRST-invariance of the boundary conditions and/or
of the out-in (one-loop) amplitude [4--8].
\vskip 0.3cm
\noindent
At about the same time, in the applications to quantum field theory
and quantum gravity, non-local boundary conditions had been 
studied mainly for operators of Dirac type 
(see, however, [9]), 
relying on the early work by Atiyah, Patodi and Singer on spectral
asymmetry and Riemannian geometry [10]. What is non-local, within
that framework, is the separation of the spectrum of a first-order
elliptic operator (the Dirac operator on the boundary) into its
positive and negative parts. This leads, in turn, to an unambiguous
identification of positive- and negative-frequency modes of the
(massive or massless) Dirac field, and half of them are set to zero 
on the bounding surface [8, 11--13].

On the other hand, non-local boundary conditions for operators of
Laplace type had already been studied quite intensively in the
literature, from at least two points of view:
\vskip 0.3cm
\noindent
(i) The rich mathematical theory of pseudo-differential
boundary value problems, where both the differential operator $P$
and the boundary operator $B$ may be replaced by 
integro-differential operators [14]. One may consider, for example,
the boundary value problem on a bounded open set $\Omega \subset
{\Re}^{n}$ with smooth boundary $\partial \Omega$:
$$
Pu=f \; \; \; \; {\rm in} \; \Omega 
\eqno (1.1)
$$
$$
Tu=\varphi \; \; \; \; {\rm at} \; 
\partial \Omega 
\eqno (1.2)
$$
where $P$ is the Laplace operator $-\bigtriangleup$ (from the 
point of view of the leading symbol, it is more convenient to
define the Laplace operator with a negative sign in front of 
all second derivatives)
and $T$ is a {\it trace operator}, e.g. $Tu=\gamma_{0}u \equiv 
[u]_{\partial \Omega}$ in the Dirichlet case, or 
$Tu=\gamma_{1}u \equiv \Bigr[u_{;N}\Bigr]_{\partial \Omega}$ in the
Neumann case, where ${ }_{;N}$ denotes the 
inward-pointing normal derivative at
$\partial \Omega$. The theory we are interested in includes both
the system
$$
{\cal A} \equiv \pmatrix{P \cr T \cr}:
C^{\infty}({\overline \Omega}) \rightarrow
C^{\infty}({\overline \Omega}) \times
C^{\infty}({\partial \Omega})
\eqno (1.3)
$$
and its solution operator, or {\it parametrix},
$$
{\cal A}^{-1} \equiv \Bigr(R \; \; K \Bigr):
C^{\infty}({\overline \Omega}) \times 
C^{\infty}({\partial \Omega}) \rightarrow
C^{\infty}({\overline \Omega}).
\eqno (1.4)
$$
With this notation, $R$ is the {\it Green operator} [14] 
solving the problem (cf (1.2))
$$
Pu=f \; \; {\rm in} \; \; \Omega \; \; {\rm and} \; \; 
Tu=0 \; \; {\rm at} \; \; \partial \Omega
\eqno (1.5)
$$
and $K$ is the {\it Poisson operator} solving the problem
(cf (1.1))
$$
Pu=0 \; \; {\rm in} \; \; \Omega \; \; {\rm and} \; \;  
Tu=\varphi \; \; {\rm at} \; \; \partial \Omega .
\eqno (1.6)
$$
The Green operator can be expressed in more detail as
$$
R=Q_{\Omega}+G
\eqno (1.7)
$$
where $Q_{\Omega}$ is the pseudo-differential operator 
(see appendix A) defined by
$$
Qf \equiv C_{n} \int {f(x)\over |x-y|^{n-2}}dx
\eqno (1.8)
$$
truncated to $\Omega$ (this implies extending $f$ by 0 on
${\Re}^{n}/\Omega$, applying $Q$, and restricting to $\Omega$),
and $G$ is a special term, called a {\it singular Green operator},
adapted to the choice of boundary conditions. Thus, to get a
general calculus, one has to consider systems of the form [14]
$$
{\cal A} \equiv \pmatrix{P_{+}+G & K \cr T & S \cr}
\eqno (1.9)
$$
for some integers $j,k,j',k'$ such that
$$
{\cal A}: C^{\infty}({\overline \Omega})^{j} \times
C^{\infty}({\partial \Omega})^{k} \rightarrow
C^{\infty}({\overline \Omega})^{j'} \times
C^{\infty}({\partial \Omega})^{k'}
\eqno (1.10)
$$
where $P$ is a pseudo-differential operator on ${\Re}^{n}$, 
$P_{+}$ is its truncation to $\Omega$, $G$ is a singular Green
operator acting in $\Omega$, $T$ is a trace operator
$T: \Omega \rightarrow {\partial \Omega}$,
$K$ is a Poisson operator
$K: {\partial \Omega} \rightarrow \Omega$,
and $S$ is a pseudo-differential operator acting on the boundary
of $\Omega$.

The trace operators we are interested in can take the form [14]
$$
T_{0}u \equiv \gamma_{0}u+T_{0}'u 
\eqno (1.11)
$$
or, instead [14], 
$$
T_{1}u \equiv \gamma_{1}u+S_{0}\gamma_{0}u+T_{1}'u .
\eqno (1.12)
$$
With this notation, one has [14]
$$
\gamma_{j}u \equiv \Bigr[({ }_{;N})^{j} u
\Bigr]_{\partial \Omega} \; \; \; \; 
j=0,1,... \; . 
\eqno (1.13)
$$
Moreover, $T_{0}'$ and $T_{1}'$ are integral operators going
from $\Omega$ to $\partial \Omega$, and the map $S_{0}$ acts
on functions on $\partial \Omega$. For example, in population
theory, one studies the condition [14]
$$
u(0)=\int_{0}^{\infty}u(t)f(t)dt
\eqno (1.14)
$$
expressing the number $u(0)$ of newborn individuals as a function
of the age profile $u(t)$. This is a special case of the
homogeneous condition, with $\Omega={\Re}_{+}$ (cf (1.11))
$$
\gamma_{0}u+T_{0}'u=0 
\eqno (1.15)
$$
with [14]
$$
-T_{0}'u=\int_{0}^{\infty}u(t)f(t)dt.
\eqno (1.16)
$$
\vskip 0.3cm
\noindent
(ii) Bose--Einstein condensation models, where integro-differential
boundary operators lead to the existence of {\it bulk states} and
{\it surface states} [15]. More precisely, given the function
$q \in L_{1}({\Re}) \cap L_{2}({\Re})$, one defines [15]
$$
q_{R}(x) \equiv {1\over 2\pi R}\sum_{l=-\infty}^{\infty}
e^{ilx /R} \int_{-\infty}^{\infty}e^{-ily/R} q(y)dy.
\eqno (1.17) 
$$
The function $q_{R}$ is, by construction, periodic with
period $2\pi R$, and tends to $q$ as $R$ tends to $\infty$. On
considering the region
$$
B_{R} \equiv \left \{ x,y: x^{2}+y^{2} \leq R^{2} \right \}
\eqno (1.18)
$$
one studies the Laplacian acting on square-integrable functions
on $B_{R}$, with non-local boundary conditions given by [15]
$$
\Bigr[u_{;N}\Bigr]_{\partial B_{R}}
+\oint_{\partial B_{R}} q_{R}(s-s')
u(R \cos(s'/R), R \sin(s'/R)) ds'=0.
\eqno (1.19)
$$
In polar coordinates, the resulting boundary value problem
reads [15]
$$
-\left({\partial^{2}u \over \partial r^{2}}
+{1\over r}{\partial \over \partial r}
+{1\over r^{2}}{\partial^{2}u \over \partial \varphi^{2}}
\right)=E u
\eqno (1.20)
$$
$$
{\partial u \over \partial r}(R,\varphi)
+R \int_{-\pi}^{\pi} q_{R}(R(\varphi-\theta)) u(R,\theta)
d \theta=0.
\eqno (1.21)
$$
For example, when the eigenvalue $E$ is positive in equation
(1.20), the corresponding eigenfunction reads [15]
$$
u_{l,E}(r,\varphi)=J_{l}(r \sqrt{E}) e^{il \varphi}
\eqno (1.22)
$$
where $J_{l}$ is the standard notation for the Bessel function
of first kind of order $l \in Z$. On denoting by $\widetilde q$
the Fourier transform of $q$, and inserting (1.22) into the 
boundary condition (1.21), one finds an equation leading,
{\it implicitly}, to the knowledge of the positive eigenvalues, i.e.
$$
\Bigr[\sqrt{E}J_{l}'(R\sqrt{E})+J_{l}(R\sqrt{E})
{\widetilde q}(l/R) \Bigr]=0.
\eqno (1.23)
$$
The solutions which decay rapidly away from the boundary are the
surface states, whereas the solutions which remain
non-negligible are called bulk states [15]. 

In the analysis of pseudo-differential boundary value
problems, one studies the heat equation for the operator $B$:
$$
\left({\partial \over \partial t}+B \right)u(t)=0 \; \; 
{\rm for} \; \; t>0
\eqno (1.24)
$$
with initial condition
$$
u(0)=u_{0}
\eqno (1.25)
$$
where $B$ is an operator acting like $P_{+}+G$ and with a domain
defined by the boundary condition $Tu=0$. Such an operator is
called a {\it realization} of $P$. One of the aims of
functional calculus is to make sense of the exponentiation 
$e^{-tB}$ under suitable assumptions on $B$. More precisely, one
wants to investigate $e^{-tB}$ so as to get detailed information
both on the {\it solutions} in terms of their data and on the
{\it kernel} of the solution operator and its trace [14]. For this
purpose, a basic tool is the analysis of the resolvent
$$
R_{\lambda} \equiv (B-\lambda I)^{-1}
\eqno (1.26)
$$
which makes it possible to study also other functions of $B$,
defined by the Cauchy integral formula
$$
f(B)={i\over 2\pi}\int_{\gamma}f(\lambda)R_{\lambda} d\lambda
\eqno (1.27)
$$
with $\gamma$ a curve in the complex plane going around the
spectrum of $B$. 

So far, we have tried to convince the general reader that there
are many good reasons for studying pseudo-differential boundary
value problems with their functional calculus on the one hand, 
and their applications to Euclidean quantum gravity on the other
hand. Now we can outline the plan of our paper, which 
is as follows. Section 2, relying on [14],
describes how to build the adjoint of a Laplacian, when non-local
boundary conditions are imposed. This scheme is then applied to
the gravitational field in section 3, with a particular choice
of integro-differential boundary conditions. A mode-by-mode form
of such boundary conditions is studied in section 4, when the
background consists of the Euclidean 4-ball. Concluding remarks
are presented in section 5, and relevant details are 
given in the appendices.
\vskip 0.3cm
\leftline {\bf 2. The adjoint with non-local boundary conditions}
\vskip 0.3cm
\noindent
In the case of the gravitational field, inspired by section 1,
we consider a scheme where the differential operator on metric
perturbations remains of Laplace type (as well as the ghost
operator), whereas the boundary conditions are of integro-differential
nature. This means that the full boundary operator, say 
$B_{ab}^{\; \; \; cd}$, may be expressed as the sum of a local
operator, say ${\widetilde B}_{ab}^{\; \; \; cd}$, obtained from
projectors and first-order differential operators [4, 5], and an
integral operator, so that the boundary conditions read
(see appendix B)
$$ \eqalignno{
\; & \Bigr[B_{ab}^{\; \; \; cd}h_{cd}(x)\Bigr]_{\partial M}
=\Bigr[{\widetilde B}_{ab}^{\; \; \; cd}
h_{cd}(x)\Bigr]_{\partial M} \cr
&+\left[\int_{M}W_{ab}^{\; \; \; cd}(x,x')
h_{cd}(x')dV' \right]_{\partial M}
&(2.1a)\cr}
$$
where $dV'$ denotes the integration measure over $M$.
We may now decide, following
DeWitt [16], that unprimed lower-case indices refer to the point
$x$ and primed lower-case indices refer to the point $x'$.
This leads to
$$
\Bigr[B_{ab}^{\; \; \; cd}h_{cd}(x)\Bigr]_{\partial M}
=\Bigr[{\widetilde B}_{ab}^{\; \; \; cd}h_{cd}(x)\Bigr]_{\partial M}
+\left[\int_{M}W_{ab}^{\; \; \; c'd'}h_{c'd'}
dV' \right]_{\partial M}
\eqno (2.1b)
$$
which is the form of the boundary conditions chosen hereafter.

Since we are concerned, for simplicity, with operators of
Laplace type in a flat four-dimensional background (all
curvature effects result then from the boundary only), it is
very important for us to understand the effect of 
integro-differential boundary conditions on such a class of
operators, motivated by section 1 and bearing in mind
Eq. (2.1b). For this purpose, following [14], we remark
that, after integration by parts, one finds the Green formula
for $P=-\bigtriangleup$, $u \in D(P)$, and $v$ in the domain
$D(P^{*})$ of the adjoint of $P$:
$$
(Pu,v)_{\Omega}=\Bigr(-\bigtriangleup u,v \Bigr)_{\Omega}
=\Bigr(u, -\bigtriangleup v \Bigr)_{\Omega}
+\Bigr({\cal U}\rho u, \rho v \Bigr)_{\partial \Omega}
\eqno (2.2)
$$
where the {\it Green matrix} [14] reads, in our case, 
$$
{\cal U}=i \pmatrix{0 & I \cr I & 0 \cr} 
\eqno (2.3)
$$
whilst $\rho$ is the (Cauchy) boundary operator, whose
action reduces to
$$
\rho u=(\gamma_{0}u, \gamma_{1}u).
\eqno (2.4)
$$
The same property (2.4) holds for $v \in D(P^{*})$. Suppose now
that the boundary conditions are expressed in the 
integro-differential form (cf (1.11)) 
$$
\Bigr[\gamma_{0}u+T_{0}'u \Bigr]_{\partial \Omega}=0.
$$
The term $\Bigr({\cal U}\rho u, \rho v \Bigr)_{\Gamma}$ in Eq. (2.2), 
which is equal to
$$
\Bigr({\cal U}\rho u, \rho v \Bigr)_{\partial \Omega}
=i (\gamma_{1}u,\gamma_{0}v)_{\partial \Omega}
+i (\gamma_{0}u,\gamma_{1}v)_{\partial \Omega} 
\eqno (2.5a)
$$
can be then re-expressed as 
$$
\Bigr({\cal U}\rho u, \rho v \Bigr)_{\partial \Omega}
=i (\gamma_{1}u,\gamma_{0}v)_{\partial \Omega}
+i(-T_{0}'u,\gamma_{1}v)_{\partial \Omega}
\eqno (2.5b)
$$
which implies that $P^{*}$, the (formal) adjoint of $P$, can
be obtained by adding to $-\bigtriangleup$ a singular Green
operator, i.e. [14]
$$
P^{*}v=-\bigtriangleup v + i T_{0}'^{*}\gamma_{1}v 
\eqno (2.6)
$$
supplemented by the local boundary condition
$$
\gamma_{0}v=0 \; \; {\rm at} \; \; \partial \Omega.
\eqno (2.7)
$$
By contrast, if the boundary conditions (cf (1.12))
$$
\Bigr[\gamma_{1}u+S_{0}\gamma_{0}u+T_{1}'u 
\Bigr]_{\partial \Omega}=0
$$
are imposed, which
modify the standard Neumann case, it is convenient to
re-express $\gamma_{1}u$, at the boundary, in the form
$$
\gamma_{1}u=-S_{0}\gamma_{0}u-T_{1}'u 
\eqno (2.8)
$$
and insert Eq. (2.8) into Eq. (2.5a). This implies that the
adjoint of $P$ now reads
$$
P^{*}v=-\bigtriangleup v + i T_{1}'^{*} \gamma_{0}v 
\eqno (2.9)
$$
subject to the local boundary condition
$$
\gamma_{1}v=0 \; \; {\rm at} \; \; \partial \Omega .
\eqno (2.10)
$$
In other words, we are discovering a property which is known
to some mathematicians, but not so familiar to physicists:
if an elliptic differential operator (here taken to be
of Laplace type) is studied with
integro-differential boundary conditions, its adjoint is a
pseudo-differential operator, subject to local 
boundary conditions.

Self-adjointness problems are properly formulated by studying 
the realization of the operator $P$ [14]. In
our case, this means adding to the Laplacian a singular Green
operator, and considering a trace operator which expresses
the integro-differential boundary conditions. More precisely,
a Dirichlet-type realization of $P=-\bigtriangleup$ 
is the operator [14]
$$
B_{D} \equiv \Bigr(-\bigtriangleup +G_{D} \Bigr)_{T_{0}}
\eqno (2.11)
$$
where (see (1.11))
$$
G_{D} \equiv K_{1}\gamma_{1}+G' 
\eqno (2.12)
$$
$$
T_{0} \equiv \gamma_{0}+T_{0}' .
\eqno (2.13)
$$
In our paper, the Poisson operators $K_{i}$, for $i=0,1$, are 
completely determined by the requirement of self-adjointness.
Indeed, the domains of $B_{D}$ and its adjoint coincide [14] if and
only if (cf (2.6) and (2.12))
$$
K_{1}=i \; T_{0}'^{*} 
\eqno (2.14)
$$
$$
G'=G'^{*} .
\eqno (2.15)
$$
Moreover, a Neumann-type realization of $P=-\bigtriangleup$
is the operator
$$
B_{N} \equiv \Bigr(-\bigtriangleup + G_{N} \Bigr)_{T_{1}} 
\eqno (2.16)
$$
where (see (1.12))
$$
G_{N} \equiv K_{0}\gamma_{0}+F' 
\eqno (2.17)
$$
$$
T_{1} \equiv \gamma_{1}+S_{0}\gamma_{0}+T_{1}' .
\eqno (2.18)
$$
Following [14], we use the notation (2.11) and (2.16) for
particular realizations of the Laplacian, but a notation
along the lines of (1.9) if we want to stress the properties
of a system in the functional calculus of the 
boundary value problem.
The domains of $B_{N}$ and its adjoint are found to coincide
[14] if and only if (cf (2.9) and (2.17))
$$
K_{0}=i \; T_{1}'^{*} 
\eqno (2.19)
$$
$$
S_{0}=-S_{0}^{*} 
\eqno (2.20)
$$
$$
F'=F'^{*} .
\eqno (2.21)
$$
\vskip 0.3cm
\leftline {\bf 3. Application to the gravitational field}
\vskip 0.3cm
\noindent
In the case of the gravitational field, our boundary operator
(2.1b) corresponds to the trace operator (2.13).
The local boundary operator ${\widetilde B}_{ab}^{\; \; \; cd}$
is taken to be the one for which the following conditions
are imposed on metric perturbations on a 3-sphere boundary
of radius $a$ [2]:
$$
\Bigr[h_{ij}\Bigr]_{\partial M}=0 
\eqno (3.1)
$$
$$
\Bigr[h_{0i}\Bigr]_{\partial M}=0 
\eqno (3.2)
$$
$$
\left[{\partial h_{00}\over \partial \tau}
+{6\over \tau}h_{00}
-{\partial \over \partial \tau}(g^{ij}h_{ij})
\right]_{\partial M}=0 
\eqno (3.3)
$$
where $\tau \in [0,a]$. Equations (3.1)--(3.3) express, to our
knowledge, the only set of local boundary conditions which are
of Dirichlet type on $h_{ij}$ and $h_{0i}$, and for which strong
ellipticity of the boundary value problem is not violated [5, 6].
Since we only want to modify the Dirichlet sector of such 
boundary conditions, which is expressed by (3.1) and (3.2), we have
to require that (see (2.1b))
$$
W_{00}^{\; \; \; c'd'}=0 \; \; \forall c',d' .
\eqno (3.4)
$$
Thus, we eventually consider the system (cf our Eq. (1.9), and
Eq. (1.6.84) in [14])
$$
{\cal A} \equiv \pmatrix{-\bigtriangleup + G \cr
T \cr} 
\eqno (3.5)
$$
where $T$ is of the type (2.13) in its $ij$ and $0i$
components, i.e.
$$
\Bigr[T_{ij}^{\; \; \; cd}h_{cd}(x)\Bigr]_{\partial M}
=\Bigr[h_{ij}(x)\Bigr]_{\partial M}
+\left[\int_{M}W_{ij}^{\; \; \; c'd'}h_{c'd'}
dV'\right]_{\partial M} 
\eqno (3.6)
$$
$$
\Bigr[T_{0i}^{\; \; \; cd}h_{cd}(x)\Bigr]_{\partial M}
=\Bigr[h_{0i}\Bigr]_{\partial M}
+\left[\int_{M}W_{0i}^{\; \; \; c'd'}h_{c'd'}
dV'\right]_{\partial M} 
\eqno (3.7)
$$
and of the type (3.3) (cf Eq. (2.18) and set $T_{1}'=0$
therein) in its normal component $h_{00}$, i.e.
$$
\Bigr[T_{00}^{\; \; \; cd}h_{cd}(x)\Bigr]_{\partial M}
=\left[{\partial h_{00}\over \partial \tau}
+{6\over \tau}h_{00}-{\partial \over \partial \tau}
(g^{ij}h_{ij})\right]_{\partial M} .
\eqno (3.8)
$$
Moreover, $-\bigtriangleup$ is the standard Laplacian on
metric perturbations in flat Euclidean 4-space, and $G$
may be viewed as the direct sum of $G_{D}$ and $G_{N}$
(cf example 1.6.16 in [14]), with
$$
G_{D}=K_{1}\gamma_{1}+G' 
\eqno (3.9)
$$
$$
G_{N}=F' 
\eqno (3.10)
$$
subject to the self-adjointness conditions
$$
\Bigr[(K_{1})_{jl}^{\; \; \; cd}h_{cd}\Bigr]_{\partial M}
=i \left[\int_{M}W_{jl}^{\; \; \; c'd'}h_{c'd'}
dV' \right]_{\partial M}^{*} 
\eqno (3.11)
$$
$$
\Bigr[(K_{1})_{0j}^{\; \; \; cd}h_{cd}\Bigr]_{\partial M}
=i \left[\int_{M}W_{0j}^{\; \; \; c'd'}h_{c'd'}
dV' \right]_{\partial M}^{*} 
\eqno (3.12)
$$
$$
\Bigr[(K_{1})_{00}^{\; \; \; cd}h_{cd}\Bigr]_{\partial M}=0
\eqno (3.13)
$$
$$
{G'}_{ab}^{\; \; \; cd}
=\left({G'}_{ab}^{\; \; \; cd}\right)^{*}
\eqno (3.14)
$$
$$
{F'}_{ab}^{\; \; \; cd}=\left({F'}_{ab}^{\; \; \; cd}\right)^{*}.
\eqno (3.15)
$$
Note that, by virtue of Eq. (3.4), the counterpart of
$T_{1}'$ (see (2.18)) vanishes in our problem (as we said after 
(3.7)), and hence $K_{0}$ vanishes as well (see (2.19)), 
so that $G_{N}$ reduces to $F'$ as we write in (3.10).
Furthermore, the condition (2.20) is satisfied
by virtue of the boundary condition (3.8).
\vskip 0.3cm
\leftline {\bf 4. Mode-by-mode equations on the Euclidean 4-ball}
\vskip 0.3cm
\noindent
At this stage it can be helpful to write down a set of equations
for perturbative modes of the gravitational field, once that the
right-hand sides of (3.6) and (3.7), and the left-hand side of (3.3),
are set to zero at the boundary. As in section 3, our background
is a portion of flat Euclidean 4-space bounded by a 3-sphere
(hereafter, $\vec x$ corresponds to local coordinates on the
3-sphere; $\vec x$ and $\tau$, altogether, correspond to the
symbol $x$ used before). This
is quite important in (one-loop) quantum cosmology [7, 8, 11--13],
and also as a first step towards more complicated field-theoretical
models. Under the above assumptions, the expansion of metric 
perturbations on a family of 3-spheres centred on the origin reads [8]
$$
h_{00}({\vec x},\tau)=\sum_{n=1}^{\infty}a_{n}(\tau)Q^{(n)}({\vec x})
\eqno (4.1)
$$
$$
h_{0i}({\vec x},\tau)=\sum_{n=2}^{\infty}\left[b_{n}(\tau)
{Q^{(n)}({\vec x})\over (n^{2}-1)}+c_{n}(\tau)S_{i}^{(n)}({\vec x})
\right]
\eqno (4.2)
$$
$$ \eqalignno{
\; & h_{ij}({\vec x},\tau)=\sum_{n=3}^{\infty}u_{n}(\tau)
\left[{Q_{\mid ij}^{(n)}({\vec x})\over (n^{2}-1)}
+{1\over 3}c_{ij}Q^{(n)}({\vec x})\right] \cr
&+\sum_{n=1}^{\infty}{1\over 3}e_{n}(\tau)c_{ij}Q^{(n)}({\vec x}) \cr
&+\sum_{n=3}^{\infty}\left[f_{n}(\tau)\Bigr(S_{i\mid j}^{(n)}({\vec x})
+S_{j \mid i}^{(n)}({\vec x})\Bigr) 
+z_{n}(\tau)G_{ij}^{(n)}({\vec x}) \right]. 
&(4.3)\cr}
$$
Here, with a standard notation, $Q^{(n)}({\vec x})$,
$S_{i}^{(n)}({\vec x})$ and $G_{ij}^{(n)}({\vec x})$ are the
scalar, transverse vector, transverse-traceless tensor 
hyperspherical harmonics on a unit 3-sphere
(with metric $c_{ij}$), respectively.
On denoting by $\mu$ a parameter with dimension [length]$^{-1}$,
by $I_{r}$ the modified Bessel function of first kind
and order $r$,
and on choosing the de Donder gauge-averaging functional,
one then finds for the transverse-traceless modes the formula [8]
$$
z_{n}(\tau)=\alpha_{n}\tau I_{n}(\mu \tau)
\eqno (4.4)
$$
whilst the vector modes read [8] 
$$
c_{2}(\tau)=\varepsilon I_{3}(\mu \tau)
\eqno (4.5)
$$
$$
c_{n}(\tau)={\widetilde \varepsilon}_{1,n}I_{n+1}(\mu \tau)
+{\widetilde \varepsilon}_{2,n}I_{n-1}(\mu \tau)
\eqno (4.6)
$$
$$
f_{n}(\tau)=\tau \left[-{{\widetilde \varepsilon}_{1,n}\over
(n+2)}I_{n+1}(\mu \tau)
+{{\widetilde \varepsilon}_{2,n}\over (n-2)}
I_{n-1}(\mu \tau)\right]
\eqno (4.7)
$$
and the scalar modes are given by [8]
$$
a_{1}(\tau)={1\over \tau}\Bigr[A_{1}I_{1}(\mu \tau)
+A_{4}I_{3}(\mu \tau)\Bigr]
\eqno (4.8)
$$
$$
e_{1}(\tau)=\tau \Bigr[3A_{1}I_{1}(\mu \tau)
-A_{4}I_{3}(\mu \tau)\Bigr]
\eqno (4.9)
$$
$$
a_{2}(\tau)={1\over \tau}\Bigr[B_{1}I_{2}(\mu \tau)
+B_{4}I_{4}(\mu \tau)\Bigr]
\eqno (4.10)
$$
$$
b_{2}(\tau)=B_{2}I_{2}(\mu \tau)-B_{4}I_{4}(\mu \tau)
\eqno (4.11)
$$
$$
e_{2}(\tau)=\tau \Bigr[3B_{1}I_{2}(\mu \tau)
-2B_{2}I_{2}(\mu \tau)\Bigr]
\eqno (4.12)
$$
$$
a_{n}(\tau)={1\over \tau}\Bigr[\rho_{1,n}I_{n}(\mu \tau)
+\rho_{3,n}I_{n-2}(\mu \tau)+\rho_{4,n}I_{n+2}(\mu \tau)\Bigr]
\eqno (4.13)
$$
$$
b_{n}(\tau)=\rho_{2,n}I_{n}(\mu \tau)
+(n+1)\rho_{3,n}I_{n-2}(\mu \tau)-(n-1)\rho_{4,n}I_{n+2}(\mu \tau)
\eqno (4.14)
$$
$$ \eqalignno{
\; & u_{n}(\tau)=\tau \left[-\rho_{2,n}I_{n}(\mu \tau)
+{(n+1)\over (n-2)}\rho_{3,n}I_{n-2}(\mu \tau) \right . \cr
& \left . +{(n-1)\over (n+2)}\rho_{4,n}I_{n+2}(\mu \tau)\right]
&(4.15)\cr}
$$
$$ \eqalignno{
\; & e_{n}(\tau)=\tau \Bigr[3\rho_{1,n}I_{n}(\mu \tau)
-2\rho_{2,n}I_{n}(\mu \tau)-\rho_{3,n}I_{n-2}(\mu \tau) \cr
&-\rho_{4,n}I_{n+2}(\mu \tau)\Bigr].
&(4.16)\cr}
$$
Now we follow the procedure outlined in the introduction in a
simpler case (see (1.20)--(1.23)), i.e. we insert the mode solutions
(4.4)--(4.16) of the eigenvalue equation for metric perturbations
into the mode-by-mode form of the boundary condition resulting
from (3.6), (3.7) and (3.3). For this purpose, it is convenient
to define
$$
\kappa_{0i} \equiv \int_{M}W_{0i}^{\; \; \; c'd'}
h_{c'd'} \; dV'
\eqno (4.17)
$$
$$
\kappa_{ij} \equiv \int_{M}W_{ij}^{\; \; \; c'd'}
h_{c'd'} \; dV' .
\eqno (4.18)
$$ 
The boundary conditions are then expressed by (3.3) jointly with
the equations
$$
\Bigr[h_{0i}+\kappa_{0i}\Bigr]_{\partial M}=0
\eqno (4.19)
$$
$$
\Bigr[h_{ij}+\kappa_{ij}\Bigr]_{\partial M}=0.
\eqno (4.20)
$$
The tensor fields $\kappa_{0i}$ and $\kappa_{ij}$, representing
the non-local contribution to the boundary conditions for a
symmetric rank-two tensor field, are themselves symmetric. They
can be therefore expanded on a family of 3-spheres centred on
the origin according to formulae entirely analogous to (4.2)
and (4.3), with the modes $\left \{ b_{n},c_{n} \right \}$ 
replaced by the modes $\left \{ {\tilde b}_{n}, {\tilde c}_{n}
\right \}$, say, and the modes $\left \{ u_{n},e_{n},f_{n},
z_{n} \right \}$ replaced by the modes $\left \{ {\tilde u}_{n},
{\tilde e}_{n}, {\tilde f}_{n}, {\tilde z}_{n} \right \}$. The
boundary conditions (3.3), (4.19) and (4.20) lead, therefore,
to the following set of equations for perturbative modes,
for all $n \geq 3$: 
$$
\left[{da_{n}\over d\tau}+{6\over \tau}a_{n}
-{1\over \tau^{2}}{de_{n}\over d\tau}\right](\tau=a)=0
\eqno (4.21)
$$
$$
[b_{n}+{\tilde b}_{n}](\tau=a)=0
\eqno (4.22)
$$
$$
[c_{n}+{\tilde c}_{n}](\tau=a)=0
\eqno (4.23)
$$
$$
[u_{n}+{\tilde u}_{n}](\tau=a)=0
\eqno (4.24)
$$
$$
[e_{n}+{\tilde e}_{n}](\tau=a)=0
\eqno (4.25)
$$
$$
[f_{n}+{\tilde f}_{n}](\tau=a)=0
\eqno (4.26)
$$
$$
[z_{n}+{\tilde z}_{n}](\tau=a)=0.
\eqno (4.27)
$$
The finite-dimensional spaces corresponding to the modes (4.5)
and (4.8)--(4.12) should, of course, be treated separately,
by requiring that
$$
\left[{da_{k}\over d\tau}+{6\over \tau}a_{k}
-{1\over \tau^{2}}{de_{k}\over d\tau}\right](\tau=a)=0
\; \; {\rm if} \; \; k=1,2
\eqno (4.28)
$$
$$
[b_{2}+{\tilde b}_{2}](\tau=a)
=[c_{2}+{\tilde c}_{2}](\tau=a)=0
\eqno (4.29)
$$
$$
[e_{2}+{\tilde e}_{2}](\tau=a)=0.
\eqno (4.30)
$$
It is now convenient to write more explicitly the tensor fields
$\kappa_{0i}$ and $\kappa_{ij}$, by using the definitions (4.17)
and (4.18) on the one hand, and the expansions analogous to (4.2)
and (4.3) on the other hand. This leads to
$$ \eqalignno{
\; & \kappa_{0i}({\vec x},\tau)=\int_{M}
\Bigr[W_{0i}^{\; \; \; 0'0'}h_{0'0'}
+2W_{0i}^{\; \; \; (0'k')}h_{0'k'}
+W_{0i}^{\; \; \; k'l'}h_{k'l'}\Bigr]dV' \cr
&=\sum_{n=2}^{\infty}\left[{\tilde b}_{n}(\tau)
{Q_{\mid i}^{(n)}({\vec x})\over (n^{2}-1)}
+{\tilde c}_{n}(\tau)S_{i}^{(n)}({\vec x})\right]
&(4.31)\cr}
$$
$$ \eqalignno{
\; & \kappa_{ij}({\vec x},\tau)=\int_{M}
\Bigr[W_{ij}^{\; \; \; 0'0'}h_{0'0'}
+2W_{ij}^{\; \; \; (0'k')}h_{0'k'}
+W_{ij}^{\; \; \; k'l'}h_{k'l'}\Bigr]dV' \cr
&=\sum_{n=3}^{\infty}{\tilde u}_{n}(\tau)
\left[{Q_{\mid ij}^{(n)}({\vec x})\over (n^{2}-1)}
+{1\over 3}c_{ij}Q^{(n)}({\vec x})\right] \cr
&+\sum_{n=1}^{\infty}{1\over 3}{\tilde e}_{n}(\tau)
c_{ij}Q^{(n)}({\vec x}) \cr
&+\sum_{n=3}^{\infty}\left[{\tilde f}_{n}(\tau)
\Bigr(S_{i \mid j}^{(n)}({\vec x})
+S_{j \mid i}^{(n)}({\vec x})\Bigr)
+{\tilde z}_{n}(\tau)G_{ij}^{(n)}({\vec x})\right].
&(4.32)\cr}
$$
One should now insert the expansions (4.1)--(4.3) into the
integrals occurring in (4.31) and (4.32), hence reading out
the modes ${\tilde b}_{n},{\tilde c}_{n},{\tilde u}_{n},
{\tilde e}_{n},{\tilde f}_{n},{\tilde z}_{n}$ for a given form
of $W_{ab}^{\; \; \; c'd'}$, chosen to be compatible with (3.4),
(3.11) and (3.12). Last, such a solution should be inserted into
the boundary conditions (4.21)--(4.30), which would be then 
re-expressed, in non-local form, uniquely in terms of modes 
for metric perturbations, which are known from (4.4)--(4.16).

For example, {\it if one assumes} that $W_{ab}^{\; \; \; c'd'}$
has distributional nature, and that suitable functions 
$\left \{ f_{1},...,f_{6} \right \}$ exist (they should be such
that the integrals we are going to build exist as Lebesgue
or Riemann integrals) for which (it is convenient to factorize
$\tau'^{-3}$ in our ansatz, because it cancels exactly a term
$\tau'^{3}$ from the integration measure $dV'$)
$$
W_{0i}^{\; \; \; 0'0'}=\tau'^{-3}f_{1}(\tau,\tau')
\delta({\vec x},{\vec x}'){\widehat \nabla}_{i}
\eqno (4.33)
$$
$$
2W_{0i}^{\; \; \; (0'k')}=\tau'^{-3}f_{2}(\tau,\tau')
\delta({\vec x},{\vec x}')\delta_{i}^{\; k}
\eqno (4.34)
$$
$$
W_{0i}^{\; \; \; k'l'}=\tau'^{-3}f_{3}(\tau,\tau')
\delta({\vec x},{\vec x}')c^{kl}{\widehat \nabla}_{i}
\eqno (4.35)
$$
where ${\widehat \nabla}_{i}$ coincides with the covariant
derivative on the boundary denoted by ${ }_{\mid i}$ so far,
the method described after Eq. (4.32) leads to, for all
$n \geq 2$,
$$ \eqalignno{
\; & {{\tilde b}_{n}(\tau)\over (n^{2}-1)}=
\int_{0}^{a}\Bigr[f_{1}(\tau,\tau')a_{n}(\tau')
+f_{2}(\tau,\tau'){b_{n}(\tau')\over (n^{2}-1)} \cr
&+f_{3}(\tau,\tau')e_{n}(\tau')\Bigr]d\tau' 
&(4.36)\cr}
$$
$$
{\tilde c}_{n}(\tau)=\int_{0}^{a}f_{2}(\tau,\tau')
c_{n}(\tau')d\tau' 
\eqno (4.37)
$$
whereas, if $n=1$, one finds
$$
f_{1}(\tau,\tau')a_{1}(\tau')+f_{3}(\tau,\tau')e_{1}(\tau')=0
\eqno (4.38)
$$
which implies the non-trivial property
$$
{f_{1}(\tau,\tau')\over f_{3}(\tau,\tau')}
=-{e_{1}(\tau')\over a_{1}(\tau')}.
\eqno (4.39)
$$
Moreover, if one assumes that
$$
W_{ij}^{\; \; \; 0'0'}=\tau'^{-3} f_{4}(\tau,\tau')
\delta({\vec x},{\vec x}')c_{ij}
\eqno (4.40)
$$
$$
2W_{ij}^{\; \; \; (0'k')}=\tau'^{-3}f_{5}(\tau,\tau')
\delta({\vec x},{\vec x}') \delta_{(i}^{\; \; k} \;
{\widehat \nabla}_{j)}
\eqno (4.41)
$$
$$
W_{ij}^{\; \; \; k'l'}=\tau'^{-3}f_{6}(\tau,\tau')
\delta({\vec x},{\vec x}') \delta_{(i}^{\; \; k}
\; \delta_{j)}^{\; \; l}
\eqno (4.42)
$$
the same method implies, for the first few modes,
$$
{\tilde e}_{1}(\tau)=\int_{0}^{a}\Bigr[3f_{4}(\tau,\tau')
a_{1}(\tau')+f_{6}(\tau,\tau')e_{1}(\tau')\Bigr]d\tau'
\eqno (4.43)
$$
$$
\int_{0}^{a}f_{5}(\tau,\tau')b_{2}(\tau')d\tau'=0
\eqno (4.44)
$$
$$
\int_{0}^{a}f_{5}(\tau,\tau')c_{2}(\tau')d\tau'=0
\eqno (4.45)
$$
and hence, up to a zero-measure set (see (4.5) and (4.11)),
$$
f_{5}(\tau,\tau')=0
\eqno (4.46)
$$
$$
{\tilde e}_{2}(\tau)=\int_{0}^{a}\Bigr[3f_{4}(\tau,\tau')
a_{2}(\tau')+f_{6}(\tau,\tau')e_{2}(\tau')\Bigr]d\tau'
\eqno (4.47)
$$
and, for all $n \geq 3$,
$$
{\tilde u}_{n}(\tau)=\int_{0}^{a}f_{6}(\tau,\tau')
u_{n}(\tau')d\tau'
\eqno (4.48)
$$
$$
{\tilde e}_{n}(\tau)=\int_{0}^{a}\Bigr[3f_{4}(\tau,\tau')
a_{n}(\tau')+f_{6}(\tau,\tau')e_{n}(\tau')\Bigr]d\tau'
\eqno (4.49)
$$
$$
{\tilde f}_{n}(\tau)=\int_{0}^{a}f_{6}(\tau,\tau')
f_{n}(\tau')d\tau'
\eqno (4.50)
$$
$$
{\tilde z}_{n}(\tau)=\int_{0}^{a}f_{6}(\tau,\tau')
z_{n}(\tau')d\tau' .
\eqno (4.51)
$$
The formulae (4.36), (4.37) and (4.48)--(4.51) for the coupled modes
should be inserted into the boundary conditions (4.22)--(4.27),
hence leading to the non-local form of the boundary conditions
resulting from the assumptions (4.33)--(4.35) and (4.40)--(4.42).
Nothing more explicit can be said unless one determines the form
of {\it all} functions $\left \{ f_{1}, ... , f_{6} \right \}$.
\vskip 0.3cm
\leftline {\bf 5. Concluding remarks}
\vskip 0.3cm
\noindent
Motivated by the recent developments in Euclidean quantum gravity
on manifolds with boundary [1--9] and Bose--Einstein condensation
models on the one hand [15], and by the progress in the functional
calculus of pseudo-differential boundary value problems on the
other hand [14], we have considered the mixed boundary conditions 
(3.3), (4.19) and (4.20) for the quantized gravitational field
on the Euclidean 4-ball. The main drawback of such boundary
conditions is, possibly, the apparent lack of an invariance 
principle leading to their derivation. On the other hand, it has
been recently proved in [5, 6] that precisely the boundary
conditions which are completely invariant under infinitesimal
diffeomorphisms on metric perturbations lead to serious technical
problems. In other words, on choosing the operator $P$ on metric
perturbations to be of Laplace type, the resulting boundary value
problem fails to be strongly elliptic [5, 6]. This is a technical
condition which requires that a unique solution should exist of
the eigenvalue equation for the leading symbol of $P$, subject
to a decay condition at infinite geodesic distance from the
boundary and to the boundary conditions of the problem [5, 6].
If there is lack of strong ellipticity, the fibre trace of the
heat-kernel diagonal acquires a non-integrable part near the
boundary [5], and hence not even the one-loop semiclassical
approximation is well defined (except, possibly, if one works
with a ``smeared" form [13] of the heat kernel for the operator $P$).

In the light of the above difficulties, the possible lack of an
invariance principle does not seem a sufficient reason for not
considering the many interesting motivations leading to the
boundary conditions (3.3), (4.19) and (4.20). In particular, the
Eqs. (4.4)--(4.16), jointly with (4.21)--(4.32), supplemented
by the self-adjointness conditions (3.11) and (3.12), lead to
what seems to be a very interesting calculational scheme for
quantum gravity on the Euclidean 4-ball (see (4.33)--(4.51)). 
At least six outstanding problems are now in sight:
\vskip 0.3cm
\noindent
(i) Is there a set of non-local symmetries in Euclidean quantum
gravity leading to our non-local set of boundary conditions
(3.3), (4.19) and (4.20) or to a suitable modification of
our scheme?
\vskip 0.3cm
\noindent
(ii) Is the resulting class of boundary value problems compatible
with strong ellipticity?
\vskip 0.3cm
\noindent
(iii) Can one build explicitly a class of bulk and surface states
in Euclidean quantum gravity with non-local boundary conditions,
inspired by the work in [15]? The idea is then to prove that,
upon inserting (4.4)--(4.16) into (4.21)--(4.32), solutions
formally analogous to the bulk and surface states of [15]
do actually exist (cf (1.20)--(1.23)).
\vskip 0.3cm
\noindent
(iv) Can one make assumptions for $W_{ab}^{\; \; \; c'd'}$ in
the boundary conditions (2.1b) which admit (4.33)--(4.35) and
(4.40)--(4.42) as a particular case? This would elucidate the
general structure of the kernel of the trace operator in
the system (3.5).
\vskip 0.3cm
\noindent
(v) Can one study heat-kernel asymptotics with non-local
boundary conditions for the gravitational field?
\vskip 0.3cm
\noindent
(vi) Can one perform a path-integral quantization, if the
non-local choice (2.1b) is made for the boundary data?
What form of boundary conditions should be imposed 
on ghost fields?
\vskip 0.3cm
If one were able to solve all such problems, an entirely new
vision would emerge in Euclidean quantum gravity, with a
non-trivial impact also on the other branches of Euclidean 
field theories. Thus, encouraging evidence exists that the
Euclidean approach continues to play a key role on the way
towards further progress in the theory of the quantized
gravitational field [8, 17].
\vskip 0.3cm
\leftline {\bf Acknowledgments}
\vskip 0.3cm
\noindent
The author is much indebted to Professor Gerd Grubb for 
correspondence, and to Ivan Avramidi and Alexander Kamenshchik
for many enjoyable years of scientific collaboration on the
problem of boundary conditions and on heat-kernel methods.
\vskip 0.3cm
\leftline {\bf Appendix A}
\vskip 0.3cm
\noindent
To be self-contained, it is important to describe the main ideas
behind the definition of pseudo-differential operators. For this
purpose, following [14], let us recall that the action of a 
differential operator of order $m$ on ${\Re}^{n}$,
$$
A(x,D_{x})=\sum_{|\alpha| \leq m} a_{\alpha}(x)D^{\alpha}
\eqno (A.1)
$$
can be expressed, with the help of Fourier transform, as
$$
A(x,D_{x})u(x)=(2\pi)^{-n}\int e^{ix {\cdot} \xi} a(x,\xi)
{\hat u}(\xi)d\xi
\eqno (A.2)
$$
where $a: (x,\xi) \rightarrow a(x,\xi)$ is the function called
{\it symbol} or characteristic polynomial, whose action is
defined by
$$
a(x,\xi) \equiv \sum_{|\alpha| \leq m}a_{\alpha}(x)\xi^{\alpha}.
\eqno (A.3)
$$
Pseudo-differential operators are obtained by considering, instead 
of a symbol of polynomial nature as in (A.3), a more general
function. More precisely, a pseudo-differential operator $P$ with
symbol $p(x,\xi)$ is the operator defined by [14]
$$
(Pu)(x) \equiv (2\pi)^{-n} \int_{{\Re}^{2n}}
e^{i(x-y){\cdot}\xi} p(x,\xi)u(y)dy d\xi.
\eqno (A.4)
$$
Strictly, the definition (A.4) holds for 
$u \in {\cal S}({\Re}^{n})$, but can be extended to a more general
class of functions, provided that the symbol $p(x,\xi)$ satisfies
suitable conditions. In particular, it is important to consider
symbols which, for some $\delta \in {\Re}$, are $C^{\infty}$
functions satisfying the inequality 
$$
\left | D_{x}^{\beta} D_{\xi}^{\alpha} p(x,\xi) \right |
\leq C_{\alpha,\beta}(x)
\left(1+|\xi|^{2} \right)^{{\delta \over 2}-
{|\alpha|\over 2}} \; \; \forall \alpha, \beta
\eqno (A.5)
$$
where $C_{\alpha,\beta}$ is a continuous function. In several
applications, one needs also the asymptotic expansion of the
symbol. For this purpose, one can assume that $p(x,\xi)$ is
{\it polyhomogeneous}, i.e. it satisfies (A.5) {\it and} has
an asymptotic expansion [14]
$$
p(x,\xi) \sim \sum_{l \in {\cal N}} p_{{\delta}-l}(x,\xi)
\eqno (A.6)
$$
where each $p_{{\delta}-l}$ is a $C^{\infty}$ function homogeneous
of degree $\delta-l$ in $\xi$ for $|\xi|>1$, and
$$
p-\sum_{l < j}p_{{\delta}-l}
$$
is a $C^{\infty}$ function satisfying the inequality (A.5) with
$\delta$ replaced by $\delta-j$, for all $j \in {\cal N}$.
\vskip 0.3cm
\leftline {\bf Appendix B}
\vskip 0.3cm
\noindent
The construction of the local part of the boundary operator in
Eq. (2.1b) deserves further comments. In other words, a rigorous
formulation of local boundary conditions for operators of Laplace
type involves vector bundles over $M$ and its boundary with their
sections, a matrix consisting of projectors and first-order
differential operators, and the boundary data. A concise description
is as follows. We consider an $m$-dimensional Riemannian manifold,
say $(M,g)$, a vector bundle $V$ over $M$, with a connection
$\nabla$, and operators of Laplace type, i.e.
$$
P \equiv -g^{ab}\nabla_{a}\nabla_{b}-E
\eqno (B.1)
$$
with $E$ an endomorphism of $V$. The operator $P$ maps smooth
sections of $V$, say $\varphi$, into smooth sections of $V$. In
the case of local boundary conditions for $P$, their general
form is [5]
$$
\pmatrix{\Pi & 0 \cr \Lambda & \II-\Pi \cr}
\pmatrix{[\varphi]_{\partial M} \cr 
[\varphi_{;N}]_{\partial M} \cr}=0
\eqno (B.2)
$$
where $\Pi$ is a self-adjoint projection operator, and $\Lambda$
is a tangential differential operator on the boundary of $M$:
$$
\Lambda \equiv (\II-\Pi)\left[{1\over 2}\Bigr(
\Gamma^{i}{\widehat \nabla}_{i}+{\widehat \nabla}_{i}
\Gamma^{i}\Bigr)+ \Sigma \right](\II-\Pi).
\eqno (B.3)
$$
With the notation used in (B.3), $\widehat \nabla$ is the induced
connection on $\partial M$, $\Gamma^{i}$ are endomorphism-valued
vector fields on the boundary, and $\Sigma$ is an endomorphism of the
vector bundle over $\partial M$ which is a copy of 
$[V]_{\partial M}$, with sections given by $[\varphi]_{\partial M}$.
$\Gamma^{i}$ and $\Sigma$ are anti-self-adjoint and self-adjoint,
respectively, and are annihilated by $\Pi$ on the left and on the
right, i.e.
$$
\Pi \; \Gamma^{i}=\Gamma^{i} \; \Pi=0
\eqno (B.4)
$$
$$
\Pi \; \Sigma =\Sigma \; \Pi =0.
\eqno (B.5)
$$
As is shown in [1, 4, 5, 7, 8], one arrives at such boundary
conditions whenever one tries to obtain gauge- and
BRST-invariant boundary conditions in quantum field theory.

In our paper, however, to avoid losing strong ellipticity of the
boundary value problem in Euclidean quantum gravity [5, 6], 
we always assume that
$$
\Gamma^{i}=0
\eqno (B.6)
$$
for all $i=1,...,m-1$.
\vskip 0.3cm
\leftline {\bf References}
\vskip 0.3cm
\noindent
\item {[1]}
Barvinsky A O 1987 {\it Phys. Lett.} {\bf 195B} 344
\item {[2]}
Luckock H C 1991 {\it J. Math. Phys.} {\bf 32} 1755
\item {[3]}
Esposito G and Kamenshchik A Yu 1995 {\it Class. Quantum Grav.}
{\bf 12} 2715
\item {[4]}
Avramidi I G, Esposito G and Kamenshchik A Yu 1996 {\it Class.
Quantum Grav.} {\bf 13} 2361
\item {[5]}
Avramidi I G and Esposito G 1997 Gauge theories on manifolds
with boundary {\it Preprint} hep-th/9710048 (to appear
in {\it Comm. Math. Phys.})
\item {[6]}
Avramidi I G and Esposito G 1998 {\it Class. Quantum Grav.} 
{\bf 15} 1141
\item {[7]}
Moss I G and Silva P J 1997 {\it Phys. Rev.} {\bf D 55} 1072
\item {[8]}
Esposito G, Kamenshchik A Yu and Pollifrone G 1997 {\it Euclidean
Quantum Gravity on Manifolds with Boundary}
({\it Fundamental Theories of Physics 85})
(Dordrecht: Kluwer).
\item {[9]}
Marachevsky V N and Vassilevich D V 1996 {\it Class. Quantum Grav.}
{\bf 13} 645
\item {[10]}
Atiyah M F, Patodi V K and Singer I M 1975 {\it Math. Proc. Camb.
Phil. Soc.} {\bf 77} 43
\item {[11]}
D'Eath P D and Halliwell J J 1987 {\it Phys. Rev.} {\bf D 35} 1100
\item {[12]}
D'Eath P D and Esposito G 1991 {\it Phys. Rev.} {\bf D 44} 1713
\item {[13]}
Esposito G 1998 {\it Dirac Operators and Spectral Geometry}
({\it Cambridge Lecture Notes in Physics 12})
(Cambridge: Cambridge University Press)
\item {[14]}
Grubb G 1996 {\it Func\-tio\-nal Cal\-cu\-lus 
of Pse\-u\-do-Dif\-fe\-ren\-ti\-al
Bo\-un\-da\-ry Pro\-ble\-ms} ({\it Pro\-gre\-ss in Ma\-the\-ma\-ti\-cs 65})
(Boston: Birkh\"{a}user)
\item {[15]}
Schr\"{o}der M 1989 {\it Rep. Math. Phys.} {\bf 27} 259
\item {[16]}
DeWitt B S 1967 {\it Phys. Rev.} {\bf 160} 1113
\item {[17]}
Gibbons G W and Hawking S W 1993 {\it Euclidean Quantum Gravity}
(Singapore: World Scientific)

\bye